\documentclass[sigconf]{acmart}

\copyrightyear{2020}
\acmYear{2020}
\setcopyright{rightsretained}
\acmConference[PPMLP'20]{2020 Workshop on Privacy-Preserving Machine Learning in Practice}{November 9, 2020}{Virtual Event, USA}
\acmBooktitle{2020 Workshop on Privacy-Preserving Machine Learning in Practice (PPMLP'20), November 9, 2020, Virtual Event, USA}\acmDOI{10.1145/3411501.3419420}
\acmISBN{978-1-4503-8088-1/20/11}

\newif\ifdraft
\newif\ifcomments
\newif\iffull
\newif\ifacademic

\draftfalse
\commentsfalse
\fullfalse
\academictrue

\usepackage{amsmath}
\usepackage{amsthm}
\usepackage{varwidth}
\usepackage[shortlabels,inline]{enumitem}
\usepackage{subcaption}
\usepackage{xspace}
\usepackage{algorithmicx}
\usepackage{algpseudocode}
\usepackage{algorithm}
\usepackage{xcolor}
\usepackage{microtype}
\usepackage{multirow}
\usepackage{balance}
\usepackage{listings}
\usepackage[binary-units]{siunitx}
    \DeclareSIUnit\bps{b/s}
\usepackage{comment}
\usepackage[noabbrev,capitalize,nameinlink]{cleveref}

\ifcomments
\usepackage[textsize=small,color=lightgray,colorinlistoftodos]{todonotes}
\else
\usepackage[disable]{todonotes}
\fi

\ifdraft
    \usepackage{fancyhdr, datetime}
\fi
\usepackage[small,compact]{titlesec}
\titleformat{\subsubsection}{\normalfont\em\bf}{\thesubsubsection}{1em}{\em}
\titlespacing*{\section}{0pt}{7pt}{3pt}
\titlespacing*{\subsection}{0pt}{4pt}{2pt}
\titlespacing*{\subsubsection}{0pt}{2pt}{0pt}

\DeclareMathAlphabet{\mathcal}{OMS}{cmsy}{m}{n}

\renewcommand{\paragraph}[1]{\noindent\textbf{#1.}\enskip}

\newcommand{\code}[1]{\texttt{#1}\xspace}

\definecolor{FGreen}{cmyk}{0.9,0.2,0.5,0.3}

\makeatletter
\newcommand*{\ie}{%
    \@ifnextchar{,}%
        {i.e.}%
        {i.e.,\@\xspace}%
}
\makeatletter
\newcommand*{\eg}{%
    \@ifnextchar{,}%
        {e.g.}%
        {e.g.,\@\xspace}%
}
\makeatletter
\newcommand*{\etc}{%
    \@ifnextchar{.}%
        {etc}%
        {etc.\@\xspace}%
}
\makeatother
\makeatletter
\newcommand*{\etal}{%
    \@ifnextchar{.}%
        {\textit{et al}}%
        {\textit{et al.}\@\xspace}%
}
\makeatother

\crefformat{section}{\S#2#1#3}

\lstdefinelanguage{sxgb}{
  sensitive = true,
  keywords=[2]{generate_client_key, encrypt_file, init_client, init_server, attest},
  keywords=[3]{DMatrix
    },
  keywordstyle=\color{purple}\bfseries,
  keywordstyle=[2]\color{red}\bfseries,
  keywordstyle=[3]\color{blue}\bfseries,
  identifierstyle=\color{black},
  sensitive=false,
  comment=[l]{\#},
  morecomment=[s]{/*}{*/},
  commentstyle=\color{gray}\ttfamily,
  stringstyle=\color{black}\ttfamily,
}

\makeatletter
\def\thickhline{%
  \noalign{\ifnum0=`}\fi\hrule \@height \thickarrayrulewidth \futurelet
   \reserved@a\@xthickhline}
\def\@xthickhline{\ifx\reserved@a\thickhline
               \vskip\doublerulesep
               \vskip-\thickarrayrulewidth
             \fi
      \ifnum0=`{\fi}}
\makeatother
\newlength{\thickarrayrulewidth}
\def\oselect{\code{oassign}}
\def\oless{\code{oless}}
\def\ogreater{\code{ogreater}}
\def\oequal{\code{oequal}}
\def\osort{\code{osort}}
\def\oarr{\code{oaccess}}

\newcommand{\eat}[1]{\ignorespaces}

\usepackage{letltxmacro}
\LetLtxMacro{\todonote}{\todo}
\usepackage{setspace}

\renewcommand{\todo}[2][]
{\todonote[size=\small,caption={#2}, #1]
{\begin{spacing}{0.5}#2\end{spacing}}}

\ifcomments
\paperwidth=\dimexpr \paperwidth + 10cm\relax
\oddsidemargin=\dimexpr\oddsidemargin + 5cm\relax
\evensidemargin=\dimexpr\evensidemargin + 5cm\relax
\marginparwidth=\dimexpr \marginparwidth + 5cm\relax
\fi

\newcommand{\sys}{Secure XGBoost\xspace}

\begin{document}
\fancyhead{}
\title{Secure collaborative training and inference for XGBoost} 

\author{Andrew Law}
\affiliation{UC Berkeley}
\authornote{Authors ordered alphabetically}
\author{Chester Leung}
\affiliation{UC Berkeley}
\author{Rishabh Poddar}
\affiliation{UC Berkeley}
\author{Raluca Ada Popa}
\affiliation{UC Berkeley}
\author{Chenyu Shi}
\affiliation{UC Berkeley}
\author{Octavian Sima}
\affiliation{UC Berkeley}
\author{Chaofan Yu}
\affiliation{Ant Financial}
\author{Xingmeng Zhang}
\affiliation{Ant Financial}
\author{Wenting Zheng}
\affiliation{UC Berkeley}

\begin{abstract}
In recent years, gradient boosted decision tree learning has proven to be an effective method of training robust models. 
Moreover, collaborative learning among multiple parties has the potential to greatly benefit all parties involved, but organizations have also encountered obstacles in sharing sensitive data due to business, regulatory, and liability concerns.

We propose \sys, a privacy-preserving system that enables multiparty training and inference of XGBoost models.
\sys protects the privacy of each party's data as well as the integrity of the computation with the help of hardware enclaves.
Crucially, \sys augments the security of the enclaves using novel \emph{data-oblivious} algorithms that prevent access side-channel attacks on enclaves induced via access pattern leakage.

\end{abstract}




\maketitle

\ifdraft
	\thispagestyle{fancy}
	\pagestyle{fancy}
\else
    \pagestyle{empty}
\fi

\section{Introduction}
\sys is a platform for secure collaborative gradient-boosted decision tree learning, based on the popular XGBoost library.
In a nutshell, multiple clients (or data owners) can \emph{collaboratively} use \sys to train an XGBoost model on their collective data in a cloud environment while preserving the privacy of their individual data. 
Even though training is done on the cloud, \sys ensures that the data of individual clients is revealed to neither the cloud environment nor other clients.
Clients collaboratively orchestrate the training pipeline remotely, and \sys guarantees that each client retains control of the computation that runs on its individual data.

At its core, \sys leverages the protection offered by \emph{secure hardware enclaves} to preserve the privacy of the data and the integrity of the computation even in the presence of a hostile cloud environment.
On top of enclaves, \sys adds a second layer of security that additionally protects the enclaves against a large class of \emph{side-channel attacks}---namely, attacks induced by access pattern leakage (see \cref{s:leakage}).
Even though the attacker cannot directly observe the data protected by the enclave, it can still infer sensitive information about the data by monitoring the enclave's memory access patterns during execution.
To prevent such leakage, we redesign the training and inference algorithms in XGBoost to be \emph{data-oblivious}, guaranteeing that the memory access patterns of enclave code does not reveal any information about sensitive data.

In implementing \sys, we strived to preserve the XGBoost API as much as possible so that our system remains easy to use for data scientists.
Our implementation has been adopted by multiple industry partners, and is available at 
\url{https://github.com/mc2-project/secure-xgboost}.

\section{Background}
\label{s:background}
\subsection{Hardware enclaves}
\label{s:background:enclaves}
Secure enclaves are a recent advance in computer processor technology that enable the creation of a secure region of memory (called an enclave) on an otherwise untrusted machine. Any data or software placed in the enclave is isolated from the rest of the system, and no other process on the same processor (not even privileged software like the OS or hypervisor) can access or tamper with that memory. Examples of secure enclave technology include Intel SGX~\cite{SGX:HASP13} and AMD Memory Encryption~\cite{AMD-ME}.

An important feature of hardware enclaves is remote attestation~\cite{sgx-attestation}, which allows a remote client system to cryptographically verify that specific software has been securely loaded into an enclave. 
As part of attestation, the enclave can also bootstrap a secure channel with the client by generating a public key and returning it with the signed report.

\subsection{Side-channel leakage}
\label{s:leakage}

A large class of known side-channel attacks on enclaves exploit data-dependent access patterns---\ie the sequence of accesses made by the executing program to disk, network, or memory. The attacker can observe the access sequence in a variety of ways: \eg cache-timing attacks~\cite{sgxcache-gotzfried, sgxcache-brasser, sgxcache-schwarz, sgxcache-moghimi, sgxattacks-hahnel:cache, sgxcache-cachequote}, branch prediction attacks~\cite{sgxattacks-lee:branches}, page monitoring~\cite{sgxattacks-xu:pagefaults, bulck-sgxattack:pagefaults}, or snooping on the memory bus~\cite{membuster}.

\paragraph{Example}
As an example, consider the code in \cref{lst:leaky} that determines the maximum of two integers using a non-oblivious if-else statement.
An attacker observing the memory addresses of accessed program instructions can identify whether $x > y$, depending on whether the code within the if-block or the else-block  gets executed.

\begin{figure}
\centering
\begin{minipage}{0.45\linewidth}
\centering
\begin{lstlisting}[showstringspaces=false]
void max(int x, int y, 
         int* z) {
  if (x > y) 
    *z = x;
  else 
    *z = y;
}
\end{lstlisting}
\captionof{figure}{Regular code}
\label{lst:leaky}
\end{minipage}%
\hspace{0.05\linewidth}
\begin{minipage}{0.45\linewidth}
\centering
\begin{lstlisting}[showstringspaces=false]
void max(int x, int y, 
         int* z) {
         
  bool cond = ogreater(x, y);
  oassign(cond, x, y, z);

}
\end{lstlisting}
\captionof{figure}{Oblivious code}
\label{lst:oblivious}
\end{minipage}
\end{figure}

\subsection{Data-obliviousness}
Oblivious computation is a type of cryptographic computation that prevents the aforementioned attacks by removing data-dependent access patterns.
Consequently, a data-oblivious enclave program prevents an attacker from inferring information about the underlying data by observing memory, disk, or network accesses.

In \sys, we design and implement data-oblivious algorithms for model training and inference. In particular, our algorithms produce an {\em identical sequence} of disk, network and memory accesses that depend only on the public information, and are {\em independent} of the input data. Hence, they provably prevent all side-channels induced by access pattern leakage.

\section{Overview}
\label{s:overview}

\subsection{System model}
\label{s:overview:architecture}
\begin{figure}
\centering
    \includegraphics[width=\columnwidth]{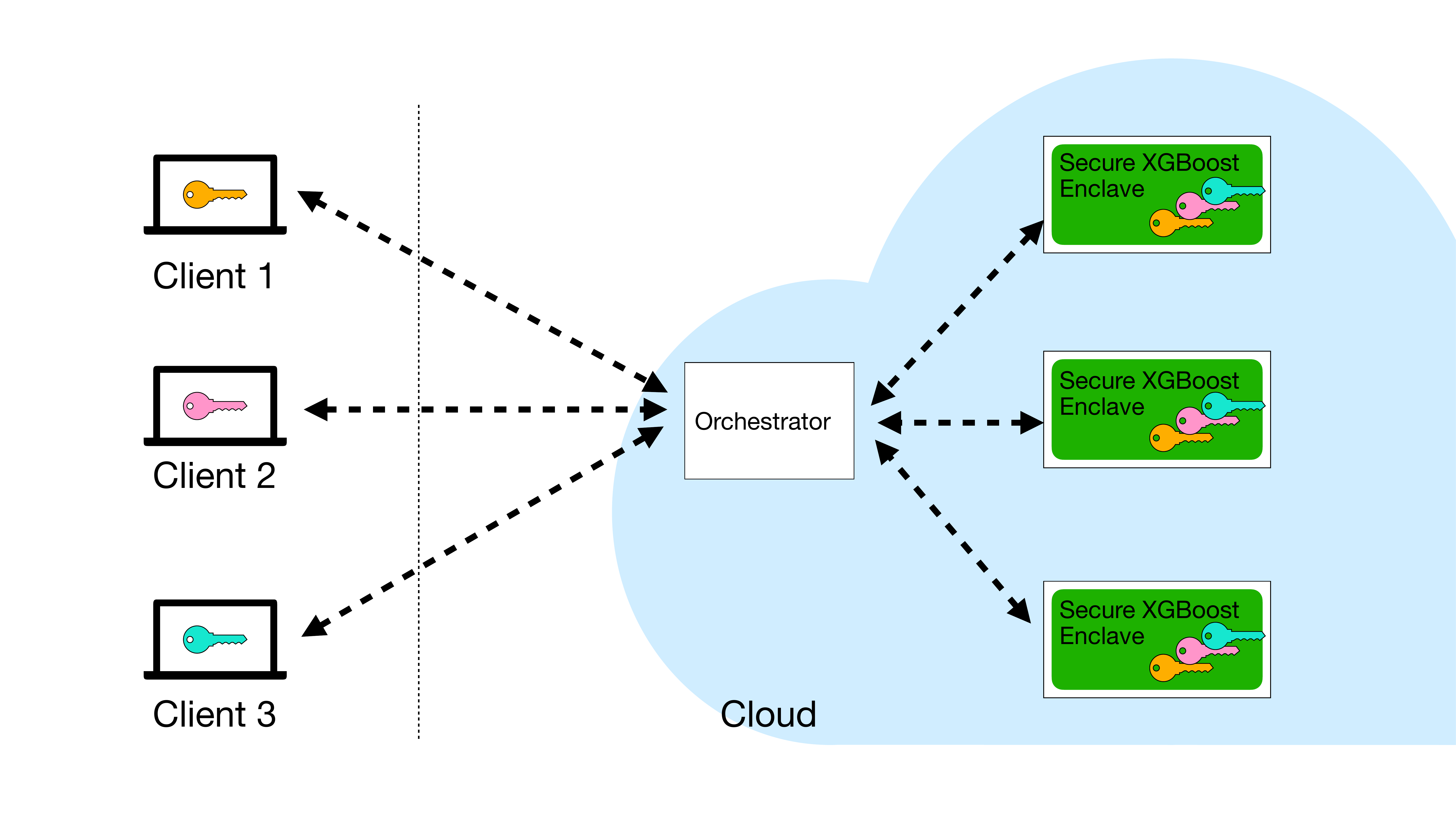}
    \caption{Parties invoke an orchestrator service at the cloud, which waits for calls from all parties before relaying the commands to the  enclave cluster. 
    Enclave inputs and outputs are always encrypted, and are decrypted only within the enclave or at client premises.}
    \label{fig:sys-arch}
\end{figure}

In this section, we describe the different entities in a \sys deployment.
The entities consist of:
\begin{enumerate*}[(i)]
\item multiple data owners (or clients) who wish to collaboratively train a model on their individual data; and
\item an untrusted cloud service that hosts the \sys platform within a cluster of machines. 
\end{enumerate*}
The general architecture of \sys is depicted in \cref{fig:sys-arch}.

\paragraph{Clients}
A client refers to a party who wants to jointly train a model with other clients.
The clients collectively execute the computation pipeline on the \sys platform by remotely invoking its APIs.

\paragraph{Cloud service with enclaves}
The cloud service consists of a cluster of virtual machines, each with hardware enclave support.
\sys distributes the computation across the cluster of hardware enclaves, which communicate with each other over TLS channels that begin and end inside the enclaves.

\label{rpc-orchestrator}
Additionally, an orchestrator service at the cloud mediates communication between clients and the \sys platform deployed within enclaves. 

\subsection{Workflow}
\label{s:overview:workflow}
The following describes an end-to-end example workflow for using \sys.
We use the term `command' to refer to a client's desired execution of a step in the computation process, \ie the APIs exposed by \sys for data loading, training, \etc.

\begin{enumerate}
    \item The clients agree on a pre-determined sequence of commands that will be executed on \sys (\cref{s:design:setup}).
    
    \item Clients attest the enclaves on the cloud (via the remote attestation procedure) to verify that the expected \sys code has been securely loaded within each enclave (\cref{s:design:attestation}). 
    
    \item Each client $C_i$ encrypts its data with a symmetric key $k_i$ and uploads it to cloud storage (\cref{s:design:data}).
    
    \item The clients submit signed commands to the orchestrator. The orchestrator aggregates all the client signatures and relays each command to \sys.
    \sys authenticates the signatures, ensuring that every client indeed issued the same command, and executes the command (\cref{s:design:execution}).
    
    \item \sys returns the results of the command (\eg an encrypted trained model, or encrypted prediction results) to the orchestrator, who relays it to the clients. The process continues until all commands have been executed.
\end{enumerate}

\section{Threat model and security guarantees}
\label{s:threatmodel}
We describe the aims and capabilities of the attackers that \sys protects against.
\subsection{Threat model for the cloud and hardware enclaves}
\label{s:threatmodel:enclaves}


The cloud service provider and the orchestrator service are untrusted. 
The trusted computing base includes the CPU package and its hardware enclave implementation, as well as our implementation of \sys.

The design of \sys is not tied to any specific hardware enclave; instead, \sys builds on top of an {\em abstract} model of hardware enclaves where the attacker controls the server’s software stack outside the enclave (including the OS), but cannot perform any attacks to glean information from inside the processor (including processor keys). 
The attacker can additionally observe the contents and access patterns of all (encrypted) pages in memory, for both data and code. We
assume that the attacker can observe the enclave’s memory access patterns at cache line granularity.

\sys provides protection against {\em all channels of attack that exploit data-dependent access patterns at cache-line granularity}, which represent a large class of known attacks on enclaves (\eg \cite{sgxcache-gotzfried, sgxcache-brasser, sgxcache-schwarz, sgxcache-moghimi, sgxattacks-hahnel:cache, sgxattacks-lee:branches, sgxattacks-xu:pagefaults, bulck-sgxattack:pagefaults, membuster}). 
Other attacks that violate our abstract enclave model---such as attacks based on timing analysis or power consumption~\cite{SGX:attack:Plundervolt,Attack:Clkscrew}, denial-of-service attacks~\cite{SGXattack:rowhammer:SGXbomb:DOS:Jang,SGX:attack:rowhammer}, or rollback attacks~\cite{Memoir:rollback} (which have complementary solutions~\cite{SGX:LCM:defense:rollback:Brandenburger:2018, SGX:ROTE:defense:rollback:Matetic:2017})---are out of scope.
Transient execution attacks (e.g., \cite{sgxattacks-foreshadow,SGX:attack:ZombieLoad,sgxattacks-sgxpectre}) are also out of scope;
these attacks violate the threat model of SGX and are typically patched promptly by the enclave vendor via microcode updates.

\subsection{Threat model for the clients}
\label{s:threatmodel:clients}
Each client expects to protect its data from the cloud service hosting the enclaves, as well as the other clients in the collaboration.
Malicious clients may collude with each other and/or the cloud service to try and learn a victim client's data.
They may also attempt to subvert the integrity of the computation by tampering with the computation steps (\ie the commands submitted for execution).
\sys protects the client data and computation in accordance with the threat model and guarantees from \cref{s:threatmodel:enclaves}.

\section{System Design}
\label{s:design}

\subsection{System setup}
\label{s:design:setup}
\sys is launched at the cloud service within enclaves.
It contains an embedded list of client names, along with the public key of a trusted certificate authority (CA), which it uses to verify a client's identity before establishing a connection with the client (described in \cref{s:design:attestation}).
A single ``master'' enclave generates a 2048-bit RSA key pair ($pk$, $sk$) and a nonce $N$. 
The public key will be used to establish a secure channel of communication with the clients, and the nonce to ensure freshness of communicated messages.

Each client $C_i$ generates a 256-bit symmetric key $k_i$.
Each client also has its own 2048-bit RSA key pair ($pk_i$, $sk_i$), along with a certificate signed by a certificate authority (CA); the CA's public key is embedded in \sys.
The clients will use the certificate to authenticate themselves to \sys.

\subsection{Client-server attestation}
\label{s:design:attestation}
Clients authenticate the \sys deployment within the enclave cluster via remote attestation (as described in \cref{s:background:enclaves}). 
More precisely, we logically arrange the enclaves in a tree topology; the enclave at the root of the tree is the ``master'' enclave.
During attestation, each client attests only the ``master'' enclave to verify that the expected \sys code has been securely loaded; in turn, each enclave in the tree (including the master) attests its children enclaves.
As part of the attestation process, the enclaves establish TLS sessions with their neighboring enclaves.
In addition, the master enclave sends the generated public key $pk$ and a nonce $N$ to the clients along with the signed attestation report.

Each client encrypts its key $k_i$ using the enclave's public key $pk$, and signs the message. It then sends the signed message to the master enclave along with its certificate.
The master enclave verifies each client's signed message, decrypts the symmetric key $k_i$, and percolates $k_i$ to all attested enclaves in the cluster, giving each enclave the ability to decrypt data belonging to the client.

\subsection{Data preparation and transfer}
\label{s:design:data}
Each client uploads its encrypted data to cloud storage; enclaves retrieve the encrypted data from storage before training.
To enable distributed data processing, each enclave must retrieve only a partition of the encrypted training data.
This requirement precludes each client from encrypting its data as a single blob.
Instead, to facilitate distributed processing of the encrypted data, clients encrypt each row in their data separately, which enables each enclave to retrieve, decrypt, and process only a subset of the rows.

Specifically, client $C_i$ encrypts each row in its data (using its symmetric key $k_i$) as follows:
\begin{center}
    $j$, \enskip $n_i$, \enskip $\code{Enc}(\code{row}_j)$, \enskip $\code{MAC}(j || n_i || \code{Enc}(\code{row}_j))$
\end{center}
Here, $j$ is the index number of the row being encrypted; $n_i$ is the total number of rows in $C_i$'s data; $\code{Enc}(\code{row}_j)$ is an AES-GCM ciphertext over the $j$-th row; and $\code{MAC}(j || n_i || \code{Enc}(\code{row}_j))$ is an AES-GCM authentication tag computed over the ciphertext, the index number $j$ and the total number of rows $n_i$.
Including $j$ and $n_i$ within the authentication tag prevents the untrusted cloud service from tampering with the data (\eg by deleting or duplicating rows).

While processing a client's data, each enclave retrieves a subset of the encrypted rows. The enclaves then communicate to ensure that they together loaded $n_i$ rows, and that all row indices from $j = 1 \ldots n_i$ were present in the retrieved data.

\subsection{Collaborative API execution}
\label{s:design:execution}

Once all clients have uploaded their data to the cloud, they collectively invoke the APIs exposed by \sys.
Each API invocation requires consensus---\sys executes an API call only if it receives the command from every client.
This ensures that no processing can be performed on a particular client's data without that client's consent.

To make an API call, each client submits a signed command to the orchestrator:
\begin{center}
\code{cmd} = <\code{seqn}, \code{func}, \code{params}>, \enskip $\code{Sign}(\code{cmd})$
\end{center}
A command contains three fields: 
\begin{enumerate*}[(i)]
    \item a sequence number \code{seqn} = $(N || \code{ctr})$ that consists of the nonce $N$ (obtained from the enclaves during attestation) concatenated with an incrementing counter;
    \item the API function \code{func} being invoked; and
    \item the function parameters \code{params}.
\end{enumerate*}
Including the sequence number ensures the freshness of the command, and prevents replay attacks on the system.
The orchestrator aggregates the signed commands and relays them to the enclave cluster.
Each enclave verifies that an identical command was submitted by every client before executing the corresponding function.

Once the function completes, \sys produces a signed response and returns it to the clients via the orchestrator:
\begin{center}
\code{resp} = <\code{seqn}, \code{result}>, \enskip $\code{Sign}(\code{resp})$
\end{center}
The response contains the sequence number of the request (to cryptographically bind the response to the request), along with the results of the function (which are potentially encrypted with the clients' keys, depending on the function that was invoked).

\section{Data-oblivious training and inference}
\label{s:obliviousness}
To prevent side-channel leakage via access patterns, we design data-oblivious algorithms for training and inference.
To implement the algorithms, we use a small set of data-oblivious primitives, based on those from prior work \cite{Ohrimenko:ObliviousML,Visor}.
In this section, we first describe the primitives, and then show their usage in our algorithms.

\subsection{Oblivious primitives}
\label{s:background:primitives}
Our oblivious primitives operate solely on registers whose contents are loaded from and stored into memory using deterministic memory accesses.
Since registers are private to the processor, any register-to-register operations cannot be observed by the attacker.

\paragraph{1) Oblivious comparisons (\oless, \ogreater, \oequal)}
These primitives can be used to obliviously compare variables, and are wrappers around the x86 \code{cmp} instruction. 

\paragraph{2) Oblivious assignment (\oselect)}
The \oselect primitive performs conditional assignments, moving a source to a destination register if a condition is true.

\paragraph{3) Oblivious sort (\osort)}
The \osort primitive obliviously sorts a size $n$ array by passing its inputs through a bitonic sorting network~\cite{BatcherSort}, which performs an identical sequence of $O(n \log^2(n))$ carefully arranged compare-and-swap operations regardless of the input array values.

\paragraph{4) Oblivious array access (\oarr)}
The \oarr primitive accesses the $i$-th element in an array without leaking $i$ itself by scanning the array at cache-line granularity while performing \oselect operations, setting the condition to true only at index $i$.

\paragraph{Example}
To show how these primitives can be used to implement higher-level data-oblivious code, \cref{lst:oblivious} depicts the data-oblivious version of the \code{max} program from \cref{lst:leaky}. In this version, all instructions are executed sequentially, without any secret-dependent branches, causing the program to have identical memory access patterns regardless of the inputs values.

\subsection{Oblivious training}
Each enclave in the cluster loads a subset of the collected data, and then uses a distributed algorithm to train the model.
In particular, we use XGBoost's histogram-based distributed algorithm (\code{hist}) for training an approximate model~\cite{xgboost,xgb:hist}, but redesign the algorithm in order to make it data-oblivious. In this algorithm, the data samples always remain distributed across all the enclave machines in the cluster, and the machines only exchange data summaries with each other. The summaries are used to construct a single tree globally and add it to the model's ensemble.

At a high level, the \code{hist} algorithm builds a tree in rounds, adding a node to the tree per round.
Given a data sample $x \in \mathbb{R}^d$, at each node the algorithm chooses a feature $j$ and a threshold $t$ according to which the data samples are partitioned (\ie if $x(j) < t$, the sample is partitioned into the left subtree, otherwise the right).
To add a node to the tree, each enclave in the cluster builds a histogram over its data for each feature; the boundaries of the bins in the histogram serve as potential splitting points for the corresponding feature. 
The algorithm combines the histograms across enclaves, and uses the aggregate statistics to find the best feature and splitting point.
Note that in the absence of data-obliviousness the algorithm reveals a large amount of information via access-pattern leakage: \eg it leaks which feature was chosen at each node in the tree, as well the complete ordering of the data samples.
We now describe the oblivious algorithm in more detail.

\begin{figure}[t]
    \centering
    \includegraphics[width=\columnwidth]{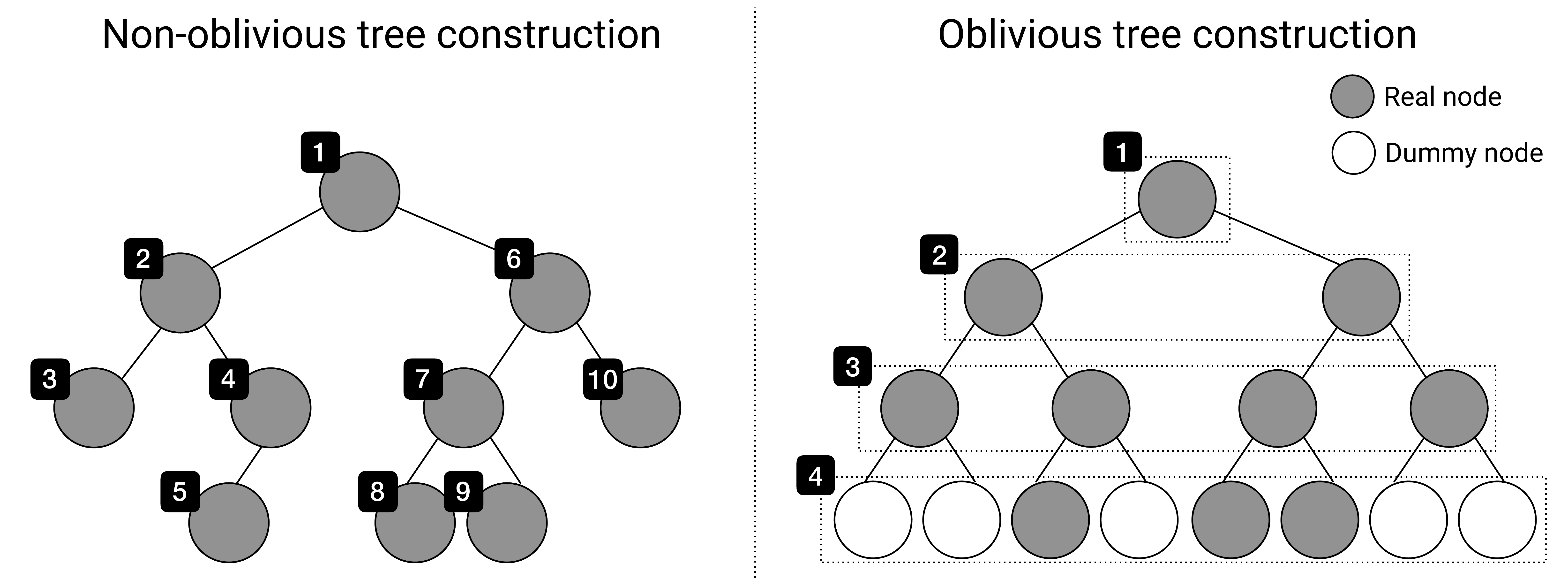}
    \caption{Illustration of oblivious training in Secure XGBoost. Numbers indicate the order in which nodes are added. Non-oblivious training adds nodes sequentially to the tree, while our algorithm constructs a full binary tree while adding nodes level-wise.}
    \label{fig:training}
\end{figure}

\paragraph{Oblivious histogram initialization}
Before a tree can be constructed, all the enclaves in the cluster first align on the boundaries of the histograms per feature. These boundaries are computed once and re-used for adding all the nodes in the tree, instead of computing new histogram boundaries per node.
\begin{enumerate}
    \item Each enclave first obliviously creates a summary $S$ of its data (one summary per feature): each element in the summary is a tuple $(y, w)$, where $y_j$ are the unique feature values in the list of data samples, and $w_j$ are the sum of the weights of the corresponding samples.
    To create the summary, the enclave sorts its samples using \code{osort}. Then, it initializes an empty array $S$ of size equal to the number of samples. Next, it scans the samples to identify unique values while maintaining a running aggregate of the weights: for each sample $\{x_i\}$ it updates $S[i]$ using \code{oselect}, either setting it to 0 (if $x_{i-1} = x_{i}$), or to the aggregated weight.
    At the end, it sorts $S$ using \code{osort} to push all 0 values to the end of the list.
    
    \item Each enclave then obliviously prunes its summary to a size $b+1$ (where $b$ is a user-defined parameter for the maximum number of bins in the histogram). The aim of the pruning operation is to select $b+1$ elements from the list with ranks $0, \frac{|S|}{b}, \frac{2|S|}{b} \ldots |S|$, where $|S|$ is the size of the summary. We do this obliviously as follows.
    First, the enclave sorts the summary using \osort. Next, it scans the sorted summary, and for each element in the summary, it selects the element (using \oselect) if its rank matches the next rank to be selected, otherwise it selects a dummy. Finally, it sorts the selected elements (which includes dummies), pushing the dummy elements to the end, and truncates the list.

    \item Next, each enclave broadcasts its summary $S$. The summaries are pairwise combined into a ``global'' summary (one summary per feature) as follows: (i)~Each pair of summaries is first merged into a single list using \osort. The tuples in the merged summary are then scanned to identify adjacent values that are duplicates; the duplicates are zeroed out using \oarr while aggregating the weights. The merged summary is then sorted using \osort to push all 0 values to the end of the list, and then truncated.
    (ii)~Next, the merged summary is pruned as before into a summary of size $b$.
\end{enumerate}
    The global summary per feature computed in this manner represents the bins of a histogram, with the constituent values in the summary as the boundaries of different bins.

\paragraph{Oblivious node addition}
The algorithm uses the feature histograms to construct a tree, adding nodes to the tree starting with the root. As nodes get added to the tree, the data gets partitioned at each node across its children. Here, we describe an oblivious subroutine for obliviously adding a node by finding the optimal split for the node, using the data samples that belong to the node.
\begin{enumerate}
    \item Each enclave computes a histogram for each feature by scanning its data samples to compute a gradient per sample, followed by updating a single bin in each histogram using \oarr combined with \oselect. The enclaves then broadcast their histograms. 
    
    \item The enclaves collectively sum up the histograms. Each enclave then computes a score function over the aggregated histogram, deterministically identifying the best feature to split by, as well as the split value.
    
    \item Finally, each enclave partitions its data based on the split value: it simply updates a marker per sample (using \oselect) that identifies which child node the sample belongs to.
\end{enumerate}

\paragraph{Level-wise oblivious tree construction}
A simple way to construct a tree is to sequentially add nodes to the tree as described above, until the entire tree is constructed. To prevent leaking information about the data or the tree: (i)~the order in which nodes are added needs to be independent of the data; and (ii)~a fixed number of nodes need to be added to the tree.
At the same time, adding nodes sequentially by repeatedly invoking the node addition subroutine above is sub-optimal for performance. This is because oblivious node addition only uses the data that belongs to the node; however, concealing which data samples belong to the node either requires accessing each sample using \oarr, or scanning all the samples while performing dummy operations for those that do not belong to the node. Both these options impact performance adversely.

We simultaneously solve all the problems above by sequentially adding entire levels to the tree, instead of individual nodes.
That is, we obliviously add all the nodes at a particular level of a tree in a \emph{single scan} of all the data samples, as follows. For each data sample, we first use \oarr to obliviously fetch the histograms of the node that the sample belongs to. We then update the histograms as described in the subroutine above, and then obliviously write back the histogram to the node using \oarr.

Note that as a result of level-wise tree construction, we always build a full binary tree (unlike the non-oblivious algorithm) and some nodes in the tree are ``dummy'' nodes. These nodes are ignored during inference.
\cref{fig:training} illustrates how nodes are added to the tree during our oblivious training routine.

\subsection{Oblivious inference}
Inference normally occurs by traversing a tree from root to leaf and comparing the feature value of each interior node with the corresponding feature in the test data instance. To obliviously evaluate an XGBoost model on a data instance, we follow \cite{Ohrimenko:ObliviousML}.  In summary, we store each layer in the tree as an array, use the \oarr primitive to obliviously select the proper node at that layer, and use the \oless primitive for comparison.

\section{Implementation}
\label{s:implementation}
In this section we describe our implementation of Secure XGBoost and discuss a few use cases in industry.

\begin{figure}[t]
\begin{lstlisting}[showstringspaces=false,language=sxgb]
import securexgboost as xgb

# Initialize client and connect to enclave cluster
xgb.init_client(user_name="user1",
				sym_key_file="key.txt",
				priv_key_file="user1.pem",
				cert_file="user1.crt")

# Client side remote attestation to authenticate enclaves
xgb.attest()

# Load the encrypted data and associate it with a user
dtrain = xgb.DMatrix({"user1": "train.enc"})
dtest = xgb.DMatrix({"user1": "test.enc"})

params = {
	"objective": "binary:logistic",
	"gamma": "0.1",
	"max_depth": "3"
}

# Train a model 
num_rounds = 5
booster = xgb.train(params, dtrain, num_rounds)

# Get encrypted predictions and decrypt them
predictions, num_preds = booster.predict(dtest)
\end{lstlisting}
\caption{Example client code in \sys. Functions highlighted in red are additions to the existing XGBoost library. Functions highlighted in blue exist in XGBoost but were modified for \sys.}
\label{api}
\end{figure}

\subsection{Implementation}
\label{s:implementation:implementation}
We implemented a prototype of \sys based on XGBoost version 0.9. Following XGBoost's implementation model, we provide a Python API on top of a core C++ library, imitating the XGBoost API as much as possible. An example of the \sys API is shown in \cref{api}. 
We used the Open Enclave SDK \cite{oesdk} to interface between the untrusted host and the enclave and to enable \sys to run agnostic of a specific hardware enclave; Mbed TLS \cite{mbedtls} for cryptography and for secure communication between enclaves; and gRPC \cite{gRPC} for client-server communication. 

Our codebase is open source and available at \url{https://github.com/mc2-project/secure-xgboost}.

\subsection{Adoption}
\label{s:implementation:adoption}
We've been fortunate enough to work with several collaborators in industry, each of whom has been using our system for a different purpose. 
Ericsson used \sys to bring together mutually distrustful network operators to collaborate in applications such as predicting cell tower hardware faults~\cite{ericsson}.
Scotiabank has been leading an effort with other Canadian banks to use \sys to fight money laundering. 
Finally, Ant Financial is using \sys in production for credit loan risk modeling. 

\begin{figure}[t]
    \centering
    \includegraphics[width=0.6\columnwidth]{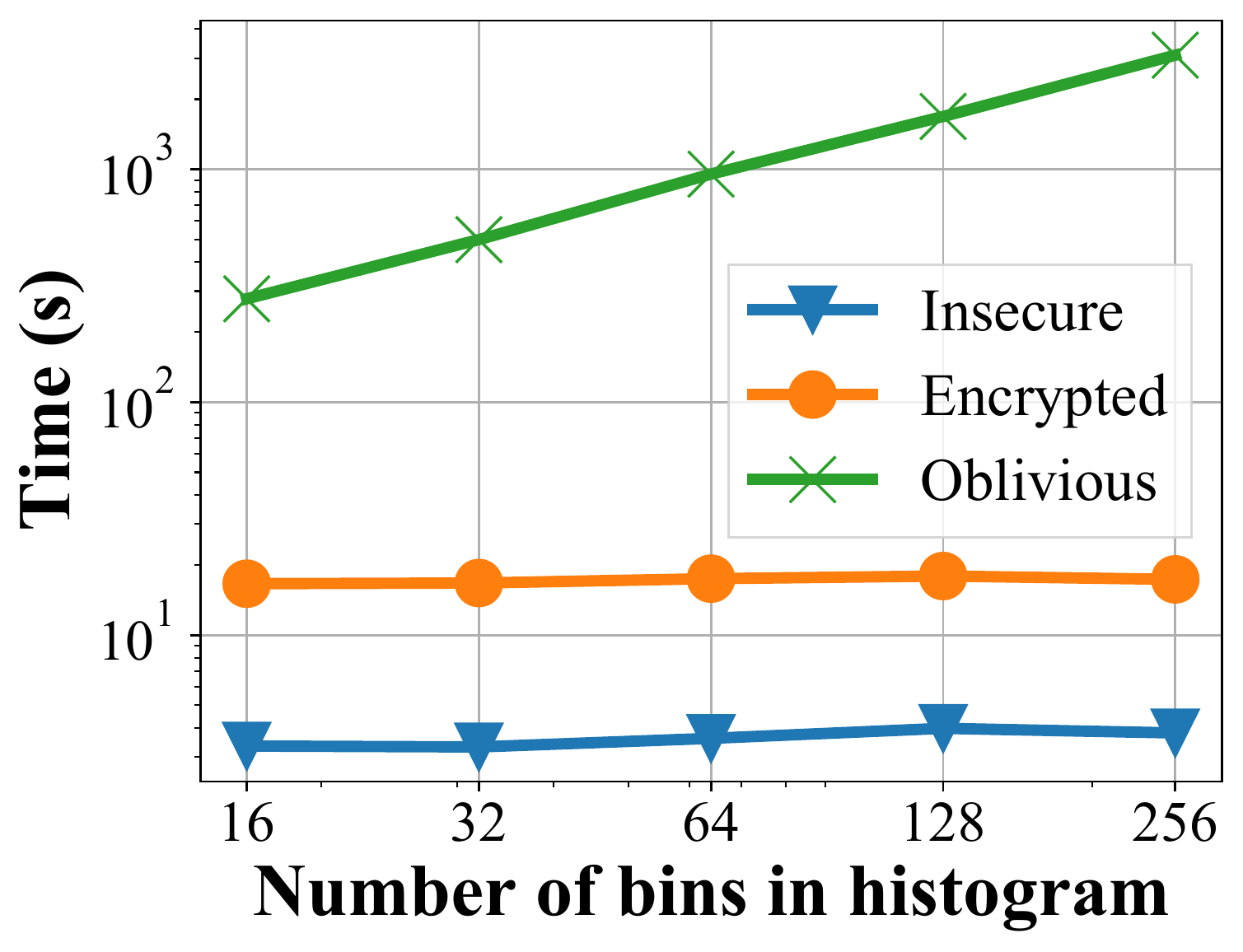}
    \caption{Evaluation comparison among the insecure baseline, and encrypted as well as oblivious \sys}
    \label{fig:overhead}
\end{figure}

\section{Evaluation results}
\label{s:evaluation}
We ran experiments on \sys using a synthetic dataset obtained from Ant Financial, consisting of $100,000$ data samples with 126 features.
Our experiments compare three systems: vanilla XGBoost; encrypted \sys (a version of \sys without obliviousness); and oblivious \sys (\sys with obliviousness enabled). 
We ran our experiments on Microsoft's Azure Confidential Computing service. 
We used DC4s\_V2 machines, which have support for Intel SGX enclaves, and are equipped with 4 vCPUs, 16 GiB of memory, and a 112 MiB enclave page cache.

\Cref{fig:overhead} shows our training results. 
In general, encrypted \sys incurs $4.5\times - 5.1\times$ overhead compared to vanilla XGBoost, which provides no security. 
Oblivious \sys incurs $16.7\times - 178.2\times$ overhead over encrypted \sys.
The main takeaway is that one has to be careful in tuning the hyperparameters by adjusting the number of bins, the number of levels per tree and the number of trees.
For example, decreasing the number of bins while increasing the number of trees could improve performance while maintaining the same accuracy.

\section{Conclusion}
\label{s:conclusion}
In this paper we proposed \sys, an oblivious distributed solution for gradient boosted decision trees using hardware enclaves. 
Our codebase is available at \url{https://github.com/mc2-project/secure-xgboost}, and we are currently working with industry collaborators to deploy our system.

\begin{acks}
This work was supported in part by the NSF CISE Expeditions Award CCF-1730628, and gifts from the Sloan Foundation, Bakar Program,  Alibaba, Amazon Web Services, Ant Group, Capital One, Ericsson, Facebook, Futurewei, Google, Intel, Microsoft, Nvidia, Scotiabank, Splunk, and VMware.
\end{acks}
\balance
\bibliographystyle{ACM-Reference-Format}
\bibliography{bib/references,bib/str,bib/conf}

\end{document}